\documentstyle [12pt,a4p,epsfig,amsmath,multicol]{article}
\begin{document}
\vspace*{0.6cm}

\begin{center} 
{\normalsize\bf Clock-transport synchronisation for neutrino 
  time-of-flight measurements}
\end{center}
\vspace*{0.6cm}
\centerline{\footnotesize J.H.Field}
\baselineskip=13pt
\centerline{\footnotesize\it D\'{e}partement de Physique Nucl\'{e}aire et 
 Corpusculaire, Universit\'{e} de Gen\`{e}ve}
\baselineskip=12pt
\centerline{\footnotesize\it 24, quai Ernest-Ansermet CH-1211Gen\`{e}ve 4. }
\centerline{\footnotesize E-mail: john.field@cern.ch}
\baselineskip=13pt
\vspace*{0.9cm}
\abstract{A method to synchronise, at the sub-nanosecond level, clocks used for neutrino
  time-of-flight measurements is proposed. Clocks situated near the neutrino source and target are compared
  with a moveable clock that is transported between them. The general-relativistic theory of the 
   procedure was tested and verified in an experiment performed by Hafele and Keating in 1972. It
  is suggested that use of such a synchronisation method may contribute to a precise test of the Sagnac 
  effect ---a measured velocity greater than $c$--- for neutrinos of the proposed LBNE beam between Fermilab
  and the Homestake mine.} 
\vspace*{0.9cm}
\normalsize\baselineskip=15pt
\setcounter{footnote}{0}
\renewcommand{\thefootnote}{\alph{footnote}}
\newline
    The aims of the present paper are twofold, firstly to present an analysis of clock transport synchronisation
     as a possible calibration method for time-of-flight measurements in two existing neutrino beams~\cite{MINOS,CNGS}
     and one proposed one~\cite{LBNE} and secondly to point out the interest of such measurements, at the level of 1ns
      accuracy, as a test of the Sagnac effect for neutrinos ---that the measured velocity is not expected
      to be exactly equal to the speed, $c$, of light in vacuum~\cite{Kuhn,JHFSagnac}. The clock transport synchronisation
       method is not proposed as a stand-alone one but rather as a cross-check of continous methods such as
       GPS Common-View GPS~\cite{GPSCV} as used for the published neutrino time-of-flight measurements or
    Two Way Satellite Time 
       Transfer~\cite{TWSTT} which have a comparable precision ($\simeq 1$ns or better) than the method proposed here.
 \par In the region of the Earth, the proper time interval, $d\tau$, of a clock is related to the
  interval of `coordinate time', $dt$, by the Schwartzschild metric equation~\cite{Schwartzs,Weinb}:
  \begin{equation} 
   d \tau = \left[1+\frac{2 \phi_{{\rm E}}}{c^2}-\frac{1}{c^2}\left(\frac{v_r^2}{1+\frac{2 \phi_{{\rm E}}}{c^2}}
      +v_{\theta}^2 +v_{\phi}^2\right)\right]^{\frac{1}{2}}dt
      \end{equation}
      where $v_r$, $v_{\theta}$ and $v_{\phi}$ are components of the velocity of the clock in
      a spherical polar coordinate system with origin at the centre of the Earth, and polar
      axis in the south-north direction, fixed in the Earth Centered Inertial (ECI) frame.
       The ECI frame is a non-rotating inertial frame instantaneously comoving with the centroid
      of the Earth. The cordinate axes of the ECI frame therefore have fixed directions relative
       to the Celestial sphere. The quantity $\phi_{{\rm E}}$ is the gravitational potential of
       the Earth,
       which, on the assumption of a spherical Earth is given as:
       $\phi_{{\rm E}} =-G M_{{\rm E}}/r$ at
       distance $r$ from the centre of the Earth, where $M_{{\rm E}}$ is the mass of the Earth.
       As can be seen by setting $\vec{v} = 0$, $\phi_{{\rm E}} = 0$ in (1), coordinate time
      is the proper 
        time of any clock at rest in the ECI frame sufficiently far from the Earth that all 
       gravitational effects of the latter may be neglected. In applying (1) it is assumed
       that the spatial variation of the gravitational fields of the Sun, Moon and other
       members of the Solar System, that give rise to tidal effects, may be neglected, as 
       well as the rotation of the Earth around the Sun that changes the angular velocity        
       of a rotating, Earth-fixed, frame relative to the ECI frame. These are the same
       approximations that are made in applications of (1) to clocks in the satellites of the
       GPS system~\cite{AshbyPT,AshbyLRR}.
      \par For a clock at rest on the Surface of the Earth $v_r = v_{\theta} = 0$  so that (1)
       simplifies to:
   \begin{equation} 
   d \tau = \left[1+\frac{1}{c^2}\left(2 \phi_{{\rm E}}
      -\Omega^2 R_{xy}^2\right)\right]^{\frac{1}{2}}dt
       \end{equation}     
   where  $R_{xy}$ is the distance of the clock from the axis of rotation of the Earth and
    $\Omega$ is the angular velocity of the latter relative to the ECI frame. Due to its rotation, the Earth has the
    form of an oblate spheroid, the surface of which is known as the geoid. Locally, in the
     oceans, the plane of the geoid is parallel to the surface of the sea. 
     \par It was shown by Cocke~\cite{Cocke} that the distortion of the Earth from a spherical
      shape modifies the external gravitational potential of the Earth in such a way that,
      to a very good approximation, the quantity $2 \phi_{{\rm E}}-\Omega^2 R_{xy}^2$ in (2)
       is constant for all points on the geoid so that identical clocks at the poles or
       on the Equator will be observed, from the ECI frame, to run at the same rate\footnote{
        Not as stated in Einstein's 1905 special relativity paper~\cite{Ein1}, slower at 
        the equator. See~\cite{HSPT} for a discussion of this point. }.
       \par If a clock moves, at fixed $r$, relative to the surface of the Earth, with speed $\vec{v}$,
        the elapsed proper time interval $\Delta \tau$
      is given by integrating (1):
       \begin{equation}
      \Delta \tau = \int\left[1+\frac{1}{c^2}\left(2 \phi_{{\rm E}}-(\vec{v}_{\Omega}+\vec{v})^2
        \right)\right]^{\frac{1}{2}} dt 
         \simeq\int\left[1+\frac{1}{c^2}\left(\phi_{{\rm E}}
       -\frac{1}{2}(\vec{v}_{\Omega}+\vec{v})^2
        \right)\right]dt 
        \end{equation}
         where $\vec{v}_{\Omega}$ is the speed, in the ECI frame, of an object, at the same position
         as the clock, but at rest relative to the Earth, 
     and where, in the last member, only terms linear in 
       $\phi_{{\rm E}}/c^2$ and quadratic in $v/c$ are retained.
        \par As suggested by Hafele~\cite{HN,HAJP} the correctness of the relation (3) was
        verified in an experiment performed by Hafele and Keating (HK)~\cite{HK} in 1972. An
       array of caesium-beam atomic clocks was flown around the Earth in commercial
       airliners in West to East (W-E) and East to West (E-W) directions. They were
       compared, before and after the flights, with reference Earth-bound clocks at the U.S.
       Naval Observatory. In agreement with prediction, time interval differences, relative to the 
      reference clocks, of 273ns for the W-E and -59ns for the E-W flights were observed~\cite{HK}.
      \par In the application of Eq.~(3) suggested, in the present work, to synchronise clocks
       for time-of-flight measurements, comparisons are made between the epochs recorded by similar
       clocks (i.e. ones running at the same frequency), one at the position of the source, and the
      other at the position of the detector, of the particles. If the clocks are placed on the geoid 
      they will be observed from the ECI frame to run at the same rate. As will be shown, the clocks 
      can be synchronised by comparing their epochs with that of a third similar clock that is
      transported between them. The single fixed reference clock of the HK experiment is therefore
       replaced by two similar clocks at different, but known, spatial locations. In order to
       make this comparison the proper time intervals $\Delta \tau_0$, $\Delta \tau$ recorded by
       the Earth-fixed and moving clocks, respectively, for the same interval, $\Delta t$, of
        coordinate time are calculated using Eq.~(3) taking account of the variation of
        the values of $\phi_{{\rm E}}$, $\vec{v}_{\Omega}$ and $\vec{v}$ along the path followed by the moveable clock.
         For the purposes of the calculation presented here it is assumed that the Earth is
         spherical; in actual experimental applications the actual shape of the geoid~\cite{AshbyLRR}
         should be taken into  account.
         \par In a practical realisation of the synchronisation method suggested here
         the clock might be equipped with a GPS receiver and associated hardware to record the precise
         position of the clock in the Earth-fixed frame at known epochs during the transport
       so that the integral in Eq.~(3) can be accurately evaluated. In the present work the dependence
         of $\Delta \tau$ on the parameters of the clock trajectory is studied within a simple model
         where only the mean speed and altitude of the transport are considered. This is sufficient to
         quantitatively demonstrate the
         predictions for $\Delta \tau$ given by different choices of transport trajectory for the three neutrino
         time-of-flight experiments that are considered. 
          \par Considering the specific case of neutrino time-of-flight measurements in the 
          CNGS neutrino beam~\cite{CNGS,OPERA} the proper time intervals recorded by reference 
         clocks at CERN ($\Delta \tau_0^{{\rm C}}$) and Gran Sasso ($\Delta \tau_0^{{\rm GS}}$), assumed
        to lie on the geoid, are given by (3), to first order in  $\phi_{{\rm E}}$ and second order in
        $\beta = v_{\Omega}/c$ as
         \begin{eqnarray}
          \Delta \tau_0 & = & \Delta \tau_0^{{\rm C}} =\int\left[1+\frac{1}{c^2}\left(\phi_{{\rm E}}^{{\rm C}}
       -\frac{1}{2}(\vec{v}_{\Omega}^{{\rm C}})^2
        \right)\right]dt   \nonumber \\
            & = & \Delta \tau_0^{{\rm GS}} = \int\left[1+\frac{1}{c^2}\left(\phi_{{\rm E}}^{{\rm GS}}
       -\frac{1}{2}(\vec{v}_{\Omega}^{{\rm GS}})^2
        \right)\right]dt
          \end{eqnarray}
           since $\phi_{{\rm E}} -(\vec{v}_{\Omega})^2/2$ is invariant on the geoid.  
        The proper time interval $\Delta \tau$ recorded by a clock transported from CERN (C) to Gran Sasso (GS)
       during the interval $\Delta t$ of coordinate time is, from (3) and (4):
       \begin{equation}
         \Delta \tau = \Delta \tau_0 + \int_{{\rm C}}^{{\rm GS}}
         \left[ \frac{G M_{{\rm E}} h}{c^2R^2}
       -\frac{1}{2c^2}[(\vec{v}_{\Omega})^2-(\vec{v}_{\Omega}^{{\rm C}})^2 +v^2+ 2\vec{v}_{\Omega}\cdot \vec{v} 
       ]\right]dt
         \end{equation}
          where $h$ is the altitude above the Earth's surface. Using a mean value approximation:
             \[ \int f(t)dt \simeq \langle f(t) \rangle \int dt \]
           to evaluate the integral in (5), and assuming transport at constant $v$, gives:
      \begin{equation}
         \frac{\Delta \tau-\Delta \tau_0}{\Delta t} \simeq  \frac{\Delta \tau-\Delta \tau_0}{\Delta \tau_0}
         \equiv \frac{\delta \tau}{\Delta \tau_0}
          = \frac{G M_{{\rm E}}\langle h \rangle}{c^2R^2}
       -\frac{1}{2c^2}[\langle (\vec{v}_{\Omega})^2\rangle-(\vec{v}_{\Omega}^{{\rm C}})^2
        +v^2+ 2\langle \vec{v}_{\Omega} \rangle \cdot \vec{v}] 
         \end{equation}
         Since the values of $\delta \tau$ in (6) are of the order of  a few nanoseconds precise modelling of
         the altitude and speed of the travelling clock in order to evaluate the integral in (5) is not required.
          For example a 5$\%$ uncertainy in the evaluation of the integral gives only $0.05$ns uncertainty in
          $\delta \tau$ when  $\delta \tau = 1$ ns.
        \par Introducing a primed, Earth-centered, Earth-fixed, polar coordinate system with north-pointing
       polar axis, $\phi' = 0$ at the prime meridian and $\lambda'$ the angle of latitude, the coordinates
       of CERN are~\cite{MJDM}: $\lambda'_{{\rm C}} = 46.05^{\circ}$, $\phi'_{{\rm C}} = 6.08^{\circ}$ 
       and of Gran Sasso are:  $\lambda'_{{\rm GS}} = 42.3^{\circ}$, $\phi'_{{\rm GS}} = 13.58^{\circ}$.
       If $\hat{\imath}'$,$\hat{\jmath}'$ and $\hat{k}'$ are unit vectors along the 
       $x'$ ($\lambda' =0$, $\phi' = 0$),
       $y'$($\lambda' =0$, $\phi' = 90^{\circ}$) and $z'$ axes of the corresponding
       right-handed Cartesian coordinate system, the spatial separation of the neutrino source at CERN
       from the OPERA detector at Gran Sasso is $d =\sqrt{(\Delta x')^2 +(\Delta y')^2+(\Delta z')^2}$
       where: 
       \begin{eqnarray}
        \Delta x' & = & R(\cos\lambda'_{{\rm GS}}\cos\phi'_{{\rm GS}}- 
        \cos\lambda'_{{\rm C}}\cos\phi'_{{\rm C}}) = 0.02936R \\          
        \Delta y' & = & R(\cos\lambda'_{{\rm GS}}\sin\phi'_{{\rm GS}}- 
        \cos\lambda'_{{\rm C}}\sin\phi'_{{\rm C}}) = 0.1003R \\
        \Delta z' & = & R(\sin\lambda'_{{\rm GS}}- \sin\lambda'_{{\rm C}}) = -0.0476R
       \end{eqnarray}
        The velocity vector $\vec{v}$ parallel to the neutrino beam is then:
        \begin{equation}
         \vec{v} = v(0.2557\hat{\imath}'+0.8734\hat{\jmath}'-0.4144\hat{k}') \equiv v \hat{v} 
         \end{equation}
         Also
        \begin{equation}
        \langle \vec{v}_{\Omega} \rangle = \Omega R(-\hat{\imath}'\langle \cos\lambda'\sin \phi'\rangle
            + \hat{\jmath}'\langle \cos\lambda'\cos \phi'\rangle)
         =  \Omega R(-0.1236\hat{\imath}'
            + 0.7047\hat{\jmath}')
       \end{equation}
        so that
         \begin{equation}
          \langle \vec{v}_{\Omega} \rangle \cdot \vec{v} = 0.5839\Omega R v
          \end{equation}
         and
          \begin{eqnarray}
          \langle (\vec{v}_{\Omega})^2\rangle & = & \Omega^2 R^2 \langle \cos^2\lambda'\rangle
             = 0.5146 \Omega^2 R^2  \\
             (\vec{v}_{\Omega}^{{\rm C}})^2 & = & \Omega^2 R^2 \cos^2\lambda'_{{\rm C}}
             = 0.4815 \Omega^2 R^2 
            \end{eqnarray}
            Substituting (12)-(14) in (6) with $\Omega R = 465$ m/s gives:
         \begin{equation}
      \delta \tau = \frac{0.00407}{v}\left[19.57 \langle h \rangle -7157-v^2-543v\right]~{\rm ns~~~CERN-CNGS} 
          \end{equation}
        with $\langle h \rangle$ in m and v in m/s. 
         Here the relation
\[ \Delta t \simeq \Delta \tau_0 = \frac{d C_{{\rm GS}}}{v} = \frac{730.53 \times 10^{3} \times 1.0022}{v} =
    \frac{732.14 \times 10^{3}}{v} \]
   is used. The straight line distance $d$ from CERN to the OPERA detector at Gran Sasso~\cite{MJDM}
      is corrected by the factor
  $C_{{\rm GS}}$ which accounts for clock transport along a Grand Circle.
   \par A similar calculation for the MINOS experiment~\cite{MINOS} in the Fermilab (FNAL) NuMI beam
         with $d = 732.7$ km, $C_{{\rm GS}}= 1.0022$, gives:
         \begin{equation}
      \delta \tau = \frac{0.00408}{v}\left[19.57 \langle h \rangle +11222-v^2+280v\right]~{\rm ns~~~FNAL-MuMI} 
          \end{equation}
      Here FNAL is assumed to be at latitude: $\lambda'_{{\rm F}} = 41.85^{\circ}$, longitude:
       $\phi'_{{\rm F}} = -88.29^{\circ}$ and the MINOS detector in the Soudan mine at 
     $\lambda'_{{\rm S}} = 47.81^{\circ}$, $\phi'_{{\rm S}} = -92.24^{\circ}$.
   \par For the proposed long-baseline FNAL-Homestake neutrino beam~\cite{LBNE} LBNE, with
      $d = 1298$ km, $C_{{\rm GS}}= 1.0069$ it is
       found that:
        \begin{equation}
      \delta \tau = \frac{0.00726}{v}\left[19.57 \langle h \rangle +4670-v^2+657v\right]~{\rm ns~~~FNAL-LBNE} 
          \end{equation}
       where the Homestake mine at Lead, South Dakota, has coordinates:
       $\lambda'_{{\rm H}} = 44.33^{\circ}$, $\phi'_{{\rm H}} = -103.83^{\circ}$~\cite{Homestake}. 
\begin{figure}[htbp]
\begin{center}\hspace*{-0.5cm}\mbox{
\epsfysize10.0cm\epsffile{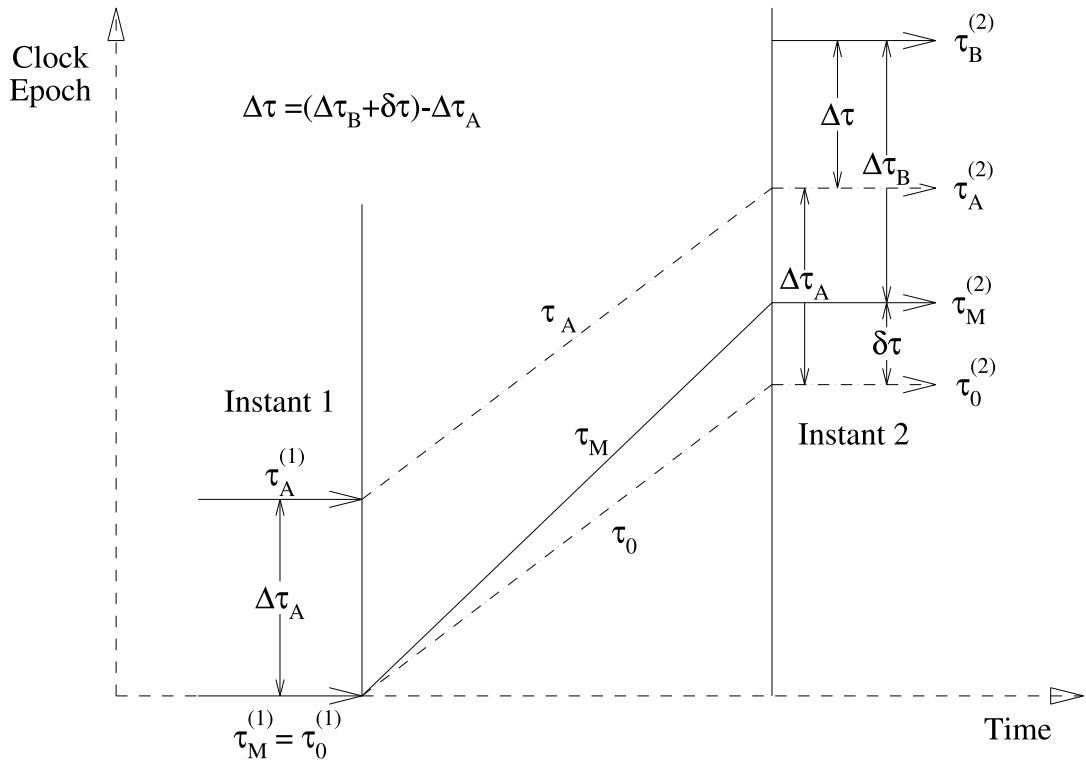}}
\caption{{\em Procedure to synchronise clocks A and B at different spatial locations on the geoid
  by comparison with a clock, M, transported at constant velocity between them. See text for discussion.}}
\label{fig-fig1}
\end{center}
\end{figure}
       \par The procedure to synchronise a clock, $A$, at the source laboratory, registering an
           epoch $\tau_{A}$ with clock $B$, at the detector laboratory, registering an
           epoch $\tau_{B}$ is as follows. Before the clock transport, the difference, $\Delta \tau_{A}$,
           between the epochs registered by clock $A$ and the moveable clock $M$ are recorded at some
           instant 1. $\Delta \tau_{A} =\tau_{A}^{(1)}-\tau_{M}^{(1)}$. After the clock transport the difference
           between the epochs of $B$ and $M$ is recorded at some instant 2:
           $\Delta \tau_{B} =\tau_{B}^{(2)}-\tau_{M}^{(2)}$. In order to synchronise clock $B$ with clock $A$
           the quantity $ \Delta \tau \equiv \Delta \tau_{B} - \Delta \tau_{A}+\delta \tau$ (see Fig.~1)
            must be subtracted from the epoch registered by clock $B$. Note that if all three clocks have the
            same frequency when at rest in the same frame of reference the differences of epochs
            $\Delta \tau_{A}$ ($\Delta \tau_{B}$) may be recorded at any time before (after)
             the transport of $M$, so that the instants 1 and 2 are arbitary.
\begin{table}
\begin{center}
\begin{tabular}{|c|c c c c c c c c c|}  \hline
 $\langle h \rangle$~(km) & 0 & 0.05 & 0.1 & 0.15 & 0.2& 0.25 & 0.3 & 0.35 & 0.36551  \\ \hline
 $v_0$~(km/h) & 305 & 283 & 260 & 234 & 205 & 171 & 129 & 63 & 0 \\
 $\delta \tau_0$~(ns) & -2.90 & -2.85 & -2.80 & -2.74 & -2.67 & -2.60 & -2.50 & -2.35 & -2.21 \\ \hline
\end{tabular} 
\caption[]{{\em  Values of $v_0$ and $\delta \tau_0$ at which $\delta \tau$ has a
     maximum value, as a function of $\langle h \rangle$, for the CERN-CGNS
   neutrino beam.}}     
\end{center}
\end{table}
            
\begin{table}
\begin{center}
\begin{tabular}{|c|c c c c c c c c|}  \hline
 $v~(km/h)$ & 10 & 50 & 100 & 200 & 400 & 600 & 800 & 1000 \\ \hline
 $\langle h \rangle$~(m)  & & & & & & & & \\ \cline{1-1} 
 0 & -12.7 & -4.4 & -3.4 & -3.0 & -2.9 & -3.1 & -3.2 & -3.4 \\
 200 & -7.0 & -3.2 & -2.8 & -2.7 & -2.8 & -3.0 & -3.2 & -3.4 \\
 400 & -1.2 & -2.1 & -2.2 & -2.4 & -2.6 & -2.9 & -3.1 & -3.3 \\
 600 & 4.5 & -0.92 & -1.7 & -2.1 & -2.5 & -2.8 & -3.0 & -3.3 \\
 800 & 10.2 & 0.22 & -1.1 & -1.8 & -2.4 & -2.7 & -3.0 & -3.2 \\
 1000 & 16.0 & 1.4 & -0.51 & -1.5 & -2.2 & -2.6 & -2.9 & -3.2 \\ \hline
\end{tabular} 
\caption[]{{\em  Values of $\delta \tau$ in ns as a function of $v$ and $\langle h \rangle$ for the CERN-CNGS
   neutrino beam.}}     
\end{center}
\end{table}

\begin{figure}[htbp]
\begin{center}\hspace*{-0.5cm}\mbox{
\epsfysize10.0cm\epsffile{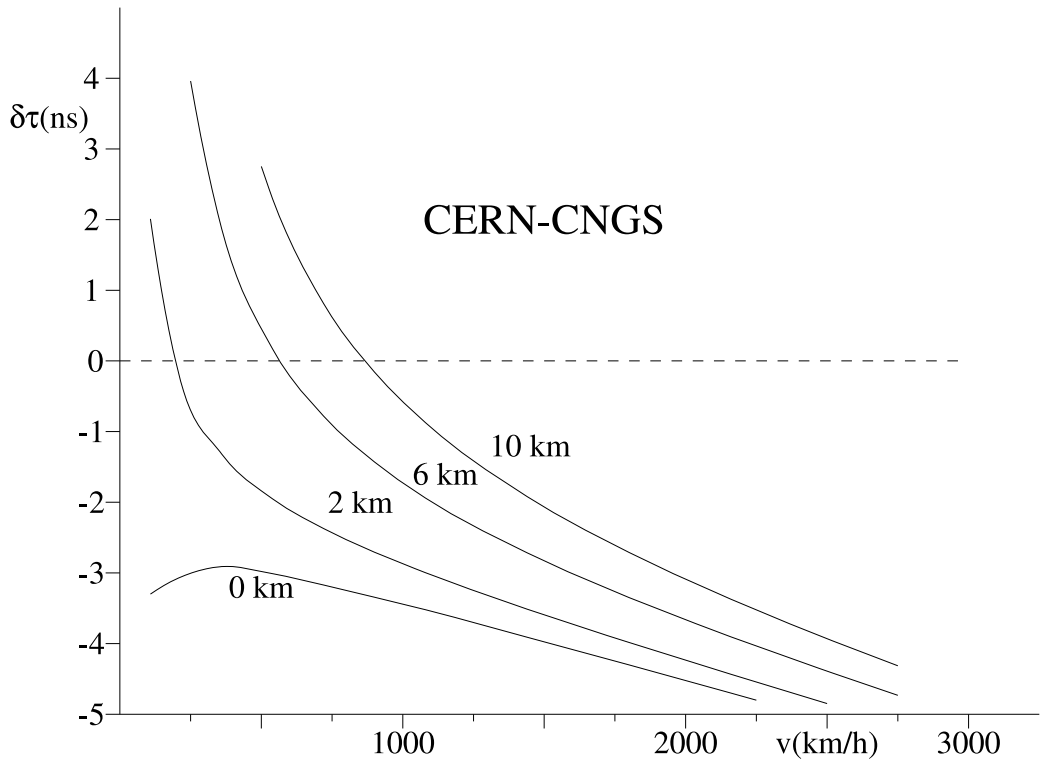}}
\caption{{\em Curves of $\delta \tau$ versus $v$ for different values of $\langle h \rangle$ for the 
  CERN-CNGS neutrino beam. Magic values of $v$ are given by the intersections of the curves with the
   dashed line $\delta \tau = 0$.}}
\label{fig-fig2}
\end{center}
\end{figure}

\begin{figure}[htbp]
\begin{center}\hspace*{-0.5cm}\mbox{
\epsfysize10.0cm\epsffile{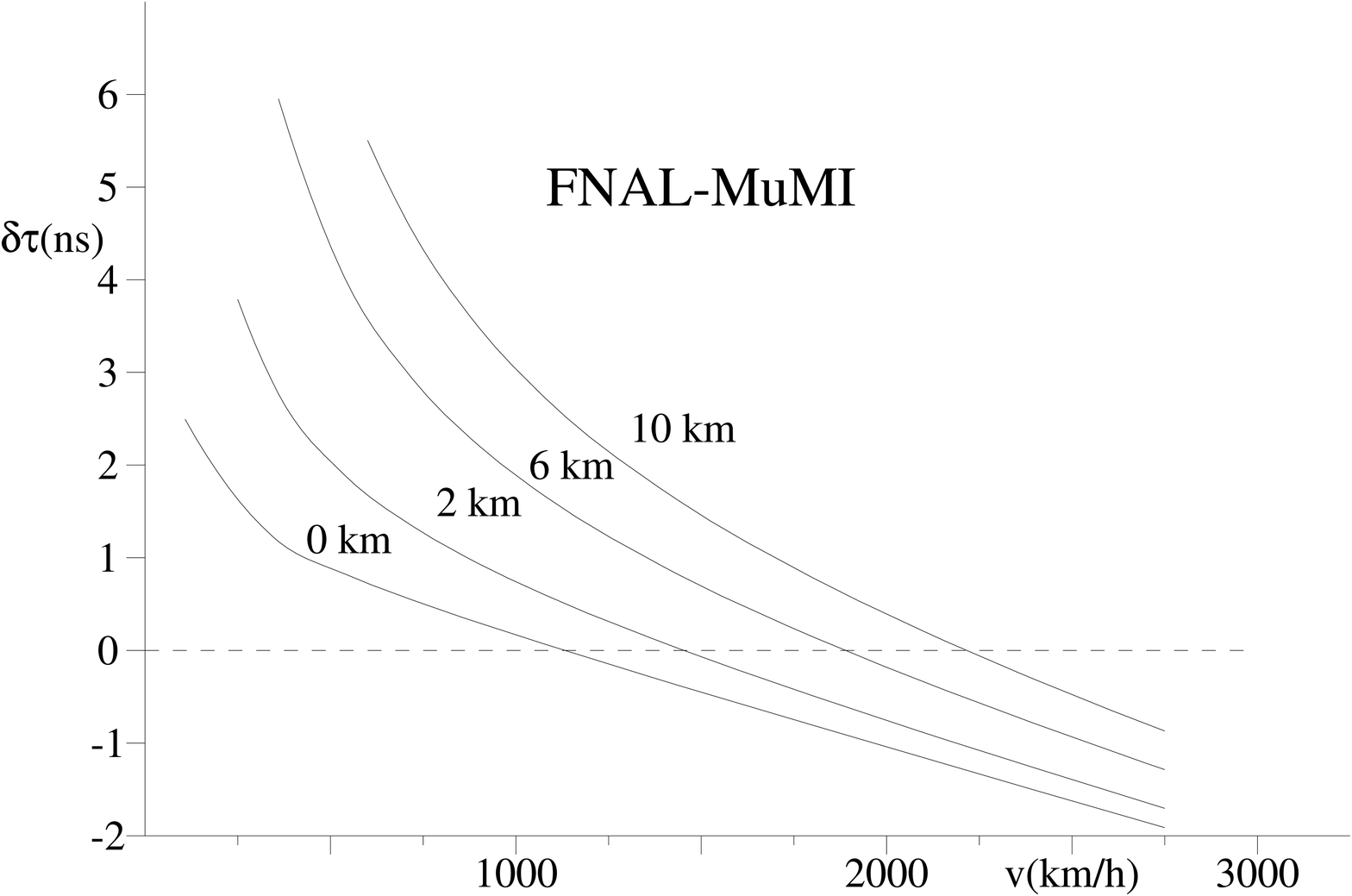}}
\caption{{\em Curves of $\delta \tau$ versus $v$ for different values of $\langle h \rangle$ for the 
  FNAL-MuMI neutrino beam.}}
\label{fig-fig3}
\end{center}
\end{figure}

\begin{figure}[htbp]
\begin{center}\hspace*{-0.5cm}\mbox{
\epsfysize10.0cm\epsffile{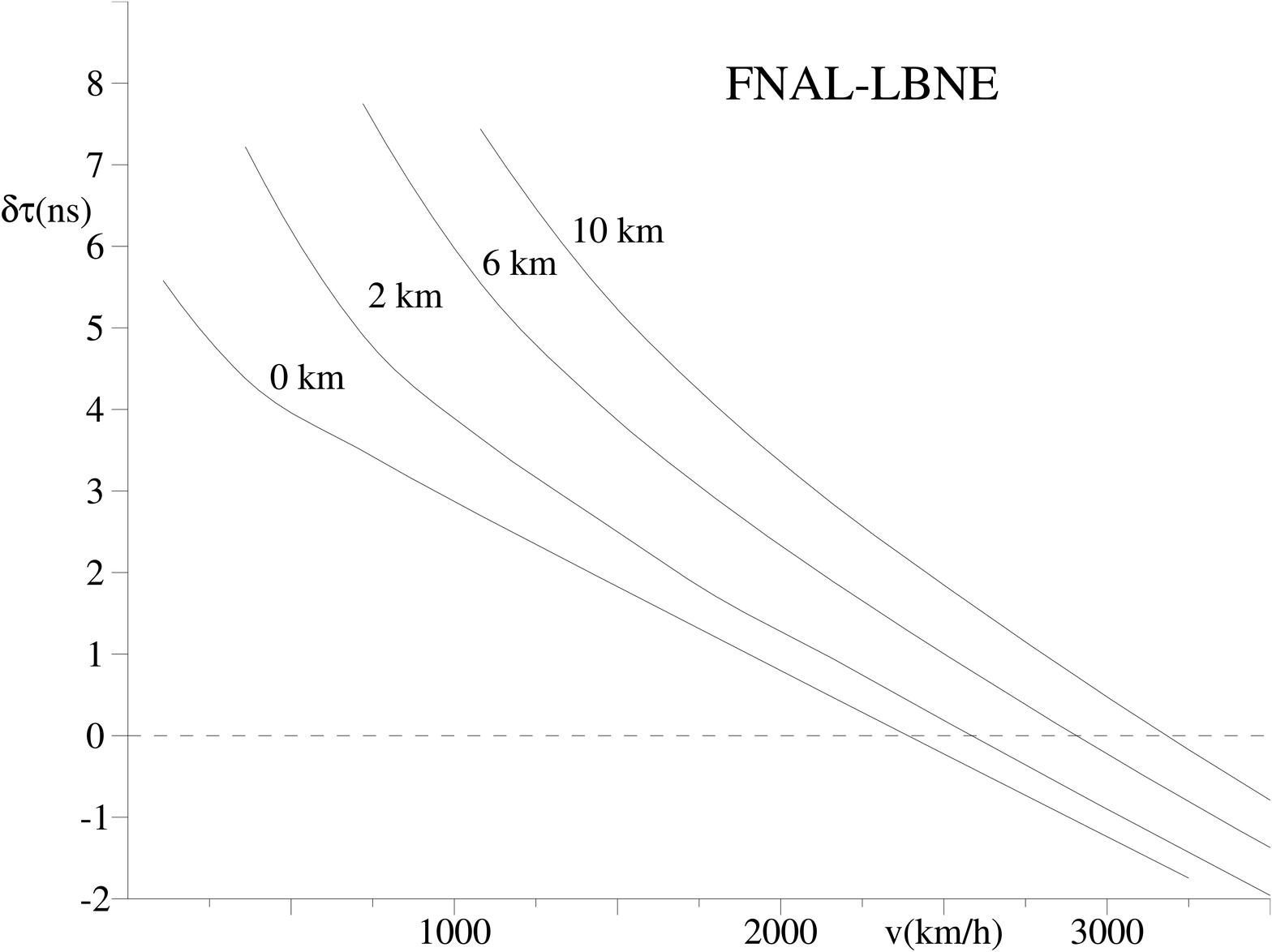}}
\caption{{\em Curves of $\delta \tau$ versus $v$ for different values of $\langle h \rangle$ for the 
   FNAL-LBNE neutrino beam.}}
\label{fig-fig4}
\end{center}
\end{figure}

       \par Curves of $\delta \tau$ versus $v$ for various values of $\langle h \rangle$ as calculated 
           by (15), (16) and (17) are shown in Figs.~2, 3 and 4 respectively. A feature of all three
         neutrino beams is the existence, for some values  of  $\langle h \rangle$, of `magic' values
         $v_{{\rm M}}$ of the velocity for which  $\delta \tau$ vanishes (given by the intersection of the
      curves with the dashed line) and so gives the possibility of
         a simple and accurate synchronisation procedure, as if the moving clock was equivalent to 
        the ground-based ones. Magic velocity values correspond to exact cancellation of the
       special-relativistic time dilation effect, which slows the rate of the moving clock as
       observed in the ECI frame, and the general-relativistic gravitational blue shift which increases
       its rate. Curves of $v_{{\rm M}}$ versus $\langle h \rangle$ for all three beams are
        shown in Fig.~5. For values of $v$: 50 km/h~$< v <$ 1000 km/h corresponding to road, rail  or air
         transport $\delta \tau$ takes, in all cases, absolute values of a few nanoseconds or less.
         \par The geometry of the CERN-CNGS beam is such that, for small values of $\langle h \rangle$,
           the curves of $\delta \tau$ versus $v$ have maximum values. The corresponding velocity
           $v_0$ and time interval $\delta \tau_0$ are given by the solution of the equation
            $d(\delta \tau)/dv = 0$ as:
             \begin{eqnarray}
             v_0 & = & \left[\langle (\vec{v}_{\Omega})^2\rangle-(\vec{v}_{\Omega}^{{\rm C}})^2
                   -\frac{2 G M_{{\rm E}}\langle h \rangle}{R^2}\right]^{\frac{1}{2}} \\
              \delta \tau_0 & = & -\frac{d C_{{\rm GC}}}{c^2}(v_0+\hat{v} \cdot \langle \vec{v}_{\Omega}\rangle)
              \end{eqnarray}
             The largest value of $\langle h \rangle$ for which a maximium value of $\delta \tau$ exists is
             $~\simeq 365$ m corresponding to $v_0 = 0$. The dependence of $v_0$ and  $\delta \tau_0$ on 
             $\langle h \rangle$ is given in Table 1. Over the range 0 m $<\langle h \rangle <$ 300 m,
             $\delta \tau_0$ varies by only $\pm 0.2$ns about an average value of 2.7ns. As shown in
             Table 2 the feature of a weak dependence of $\delta \tau$ on both $v$ and $\langle h \rangle$
             persists for  365 m $ < \langle h \rangle < $ 1000 m where the curves are monotonic. For example,
             for $v = $500 km/h and  $\langle h \rangle =$ 500m one percent uncertainties in $v$ and
              $\langle h \rangle$ give only $\simeq 0.24~\%$ and $\simeq 0.09~\%$ uncertainties, respectively
            in  $\delta \tau$, suggesting that this quantity may be determined with a precision of one tenth
            of a nanosecond or better. In order to achieve this precision over the typical beam length of
             $\simeq 1000$ km, with 100 km/h~$< v <$ 1000 km/h requires a frequency stability of the clock
             in the range $3\times 10^{-14}$---$3\times 10^{-13}$/h, compatible with the typical frequency
             stability of a precision atomic clock of 1ns/day. 
\begin{figure}[htbp]
\begin{center}\hspace*{-0.5cm}\mbox{
\epsfysize10.0cm\epsffile{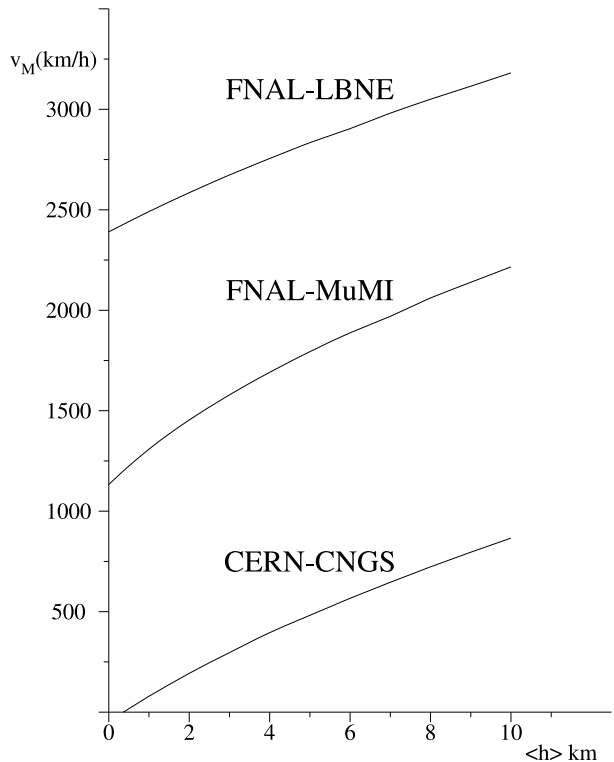}}
\caption{{\em Magic velocity values as a function of $\langle h \rangle$ for different neutrino beams.}}
\label{fig-fig5}
\end{center}
\end{figure}
            \par Since the Great Circle route from CERN to Gran Sasso passes over the Alps a possible strategy to
            keep $h$ below 1 km to take advantage of the almost constant value of  $\delta \tau$ as a function
       of $v$ and $\langle h \rangle$, as shown in Table 2, is to transport the clocks by air above the Rhone
           to the Mediterranean, then by a Grand Circle route to Gran Sasso. An alternative strategy  is a 
         Grand Circle route directly over the Alps at an altitude of 5-10km at or near the magic value of
          $v$, again enabling a precise synchronisation of the local clocks at CERN and Gran Sasso.
          \par For the FNAL neutrino beams:
            \begin{eqnarray}
          \langle (\vec{v}_{\Omega})^2\rangle-(\vec{v}_{\Omega}^{{\rm F}})^2 & = & -11.22\times 10^3
           ~~~~{\rm MuMI} \nonumber \\
          & = & -4.67\times 10^3
           ~~~~{\rm LBNE} \nonumber 
              \end{eqnarray}
           Inspection of Eq.~(18) shows that no real value of $v_0$ exists for these beams;
           the curves of  $\delta \tau$ versus $v$ are therefore monotonic for all values of $\langle h \rangle$.
            As can be seen in Figs.~3 and 4 magic values of $v$ are obtainable by conventional air
           transport at speeds of 1000-2000 km/h for the MuMI beam but not for LBNE that requires speeds in
           excess of 2000 km/h. 
           \par It is interesting to compare the uncertainty in the clock synchronisation constant $\delta \tau$
             with the variation of neutrino times-of-flight due to the Sagnac 
            effect~\cite{Kuhn,JHFSagnac,Sagnac,Post},
              i.e. the effect of the motion of the neutrino target, due to the Earth's rotation, on the observed
             time-of-flight. For neutrinos with energy much greater than their mass, the relative speed $c_r$ of
             the neutrino and the target is, at lowest order in the velocity $\vec{v}_{\Omega}^{{\rm T}}$ of the
             target, given by~\cite{Kuhn,JHFSagnac}:
             \begin{equation}
               c_r = c -\hat{v}\cdot\vec{v}_{\Omega}^{{\rm T}}
             \end{equation}
               where $\hat{v}$ is a unit vector in the direction from the neutrino source to the target.
              For the three neutrino beams considered above it is found that:
           \begin{eqnarray}
          \frac{c_r-c}{c} & = & -9.05 \times 10^{-7}~~~~{\rm CNGS} \nonumber \\
              & = & 4.65  \times 10^{-7}~~~~{\rm MuMI} \nonumber \\
             & = &  1.10  \times 10^{-6}~~~~{\rm LBNE} \nonumber 
              \end{eqnarray}
      The corresponding differences of time-of-flight compared with propagation at speed $c$ are:
      2.2 ns~\cite{Kuhn}, -1.14 ns and -4.74 ns respectively, to be compared with  the possible uncertainty in
        $\delta \tau$ of the order of a nanosecond or less. Especially the proposed FNAL-LBNE beam~\cite{LBNE}
        is well adapted to a precise test of the Sagnac effect for neutrinos. 
        \par In the case of four experiments which have published neutrino time-of-flight measurements:
          MINOS~\cite{MINOS}, OPERA\cite{OPERA}, ICARUS\cite{ICARUS} and LVD~\cite{LVD} the dominant
          systematic errors were related to local measurements of the times of production
         and detection of the neutrinos rather than errors in the synchronisation of the
         precision master clocks near to the source and target. In the LVD measuremet that
            has the smallest quoted systematic uncertainty of 3.3 ns on the  time-of-flight 
          measurement, only 1 ns is assigned to GPS synchronisation. In order to be sensitive 
         to the Sagnac effect more accurate local timimg of neutrino production and detection
         events than hitherto achieved is required in addition to synchronisation uncertainty 
         at the 1 ns level or better.
 \par {\bf Acknowledgements}
  \newline The author would like to thank two anonymous referees for pointing out several shortcomings
     in a previous version of this paper and for suggestions to improve its clarity.                                                            

\end{document}